\documentclass[conference]{IEEEtran}
\IEEEoverridecommandlockouts
\usepackage{amsmath,amssymb,amsfonts}
\usepackage{algorithmic}
\usepackage{graphicx}
\usepackage{textcomp}
\usepackage{xcolor}
\usepackage{listings}
\usepackage[hyphens]{url}

\def\BibTeX{{\rm B\kern-.05em{\sc i\kern-.025em b}\kern-.08em
    T\kern-.1667em\lower.7ex\hbox{E}\kern-.125emX}}

\usepackage{tabularx,booktabs}
\newcolumntype{C}{>{\centering\arraybackslash}X} 
\usepackage{lipsum} 
\usepackage{caption}
\usepackage[numbers]{natbib}
\usepackage{threeparttable}

\usepackage{hyperref}

\hypersetup{
    colorlinks=true,
    linkcolor=black,   
    citecolor=black,   
    urlcolor=black     
}

\begin{document}

\title{Static Application Security Testing of Consensus-Critical Code in the Cosmos Network}

\author{\IEEEauthorblockN{Jasper Surmont}
\IEEEauthorblockA{\textit{KU Leuven} \\
Leuven, Belgium \\
jasper.surmont@student.kuleuven.be}
\and
\IEEEauthorblockN{Weihong Wang}
\IEEEauthorblockA{\textit{imec-DistriNet, KU Leuven} \\
Leuven, Belgium \\
weihong.wang@kuleuven.be}
\and
\IEEEauthorblockN{Tom Van Cutsem}
\IEEEauthorblockA{\textit{imec-DistriNet, KU Leuven} \\
Leuven, Belgium \\
tom.vancutsem@kuleuven.be}
}

\maketitle

\begin{abstract}

Blockchains require deterministic execution in order to reach consensus. This is often guaranteed in languages designed to write smart contracts, such as Solidity. Application-specific blockchains or ``appchains'' allow the blockchain application logic to be written using general-purpose programming languages, giving developers more flexibility but also additional responsibilities. In particular, developers must ensure that their blockchain application logic does not contain any sources of non-determinism. Any source of non-determinism may be a potential source of vulnerabilities.

This paper focuses on the use of Static Application Security Testing (SAST) tools to detect such sources of non-determinism at development time. We focus on Cosmos, a prominent open-source project that lets developers build interconnected networks of application-specific blockchains. Cosmos provides a Software Development Kit (SDK) that allows these chains to be implemented in the Go programming language. We create a corpus of 11 representative Cosmos-based appchains to analyze for sources of non-determinism in Go.

As part of our study, we identified \emph{cosmos-sdk-codeql}, a set of CodeQL code analysis rules for Cosmos applications. We find that these rules generate many false positives and propose a refactored set of rules that more precisely detects sources of non-determinism only in code that runs as part of the blockchain logic. We demonstrate a significant increase in the precision of the rules, making the SAST tool more effective and hence potentially contributing to enhanced security for Cosmos-based blockchains.
\end{abstract}

\begin{IEEEkeywords}
Application-specific blockchain, Cosmos, Static application security testing
\end{IEEEkeywords}

\section{Introduction}

In recent years, blockchain technology has revolutionized several industries, offering new models for secure and decentralized data management. As organizations continue to explore the potential of these distributed ledgers, the need for secure blockchain software development practices becomes increasingly essential. Static Application Security Testing (SAST) tools are an important part of any secure software development lifecycle (SDLC). One of the crucial benefits of SAST tools is their ability to detect vulnerabilities early in the development process, which enables developers to address issues at an early stage, reducing cost and effort. 

The significance of SAST tools in the context of blockchain applications is even higher, and this importance can be attributed to three key factors. Firstly, in blockchains, code replication and distribution across the network make upgrades a challenging task, requiring careful coordination throughout the network. Secondly, given that blockchains handle valuable digital assets, undetected vulnerabilities can have serious implications, leading to significant financial losses. Finally, the execution of blockchain code occurs within a potentially malicious context, as adversaries can strategically craft and submit malicious payloads through transactions on the chain.

There are two approaches to creating blockchain-based applications. One approach, popularized by Ethereum~\cite{buterin:2014a}, is to develop so-called ``smart contract'' logic in a custom language (e.g. Solidity) that executes in a virtual machine as part of an application-agnostic blockchain. The other approach, popularized by projects such as Cosmos and Polkadot, is to provide developers with the tools needed to easily build their own ``application-specific'' blockchains (also called \emph{appchains}). This gives developers more flexibility in customizing the blockchain to their application needs. Appchains can form interconnected networks of blockchains, providing increased security as well as the opportunity to let apps communicate with each other. This approach has proven to be very successful with the Cosmos network, for instance, managing more than 55 billion USD in digital assets\footnote{Taken from \url{https://cosmos.network/} on 6 June 2023}.

Building appchains requires developers to invest more effort compared to solely focusing on smart contract development since they are required to implement a broader range of functionalities like the initialization of the network, light clients, and all the necessary interfaces. Frameworks like the Cosmos SDK\footnote{https://docs.cosmos.network/main} and Substrate\footnote{https://substrate.io/} play an essential role in facilitating the creation of appchains, as they allow the developers to focus on the application, rather than the blockchain internals and other functionalities.  The complexity and added responsibilities associated with appchains raise the risk of potential vulnerabilities. In particular, it is important that the blockchain can reach a consensus. A blockchain that fails to reach consensus might halt, fork, or perform other unexpected results. These outcomes can be exploited by attackers to target specific nodes, gain more power in the network, or disrupt the availability altogether. One of the primary causes of a blockchain not reaching consensus is non-determinism. Domain-specific languages such as Solidity for Ethereum limit the programs that can be written to be deterministic by design. However, in the context of an appchain where general-purpose languages are common, there are many unexpected sources of non-determinism that might put the security of the blockchain at risk, such as the use of system time. 

Because these vulnerabilities are not always straightforward to detect manually, research has been conducted to enable automated checks for such issues. \citeauthor{hyperledger-fabric-tool} \cite{hyperledger-fabric-tool} developed a SAST tool called \textit{Chaincode Analyzer} that is able to scan Hyperledger Fabric's \textit{chaincode} for common vulnerabilities, which was able to find 20 possible vulnerabilities in 15 chaincodes. \citeauthor{information-flow-analysis-nondeterminism} \cite{information-flow-analysis-nondeterminism} created the tool GoLiSa, which can detect the flow of non-deterministic operations to the store of the blockchain in Hyperledger Fabric and the Cosmos SDK. GoLiSa was evaluated on 651 Hyperledger Fabric chaincodes, and was able to detect 124 possible non-determinism-related vulnerabilities with a precision rate of 63.59\%.

\subsection{Problem statement}

For the specific case of Cosmos SDK, two additional tools have been developed.  Firstly, the Cosmos team extended the capabilities of Gosec \cite{gosec}, a Go security checker, by introducing Cosmos SDK-specific rules \cite{gosec-cosmos}. These rules enable the detection of vulnerabilities such as non-determinism or the use of unsafe packages. Secondly, CodeQL was utilized by \citeauthor{cosmos-sdk-codeql}, who created a query pack named \textit{cosmos-sdk-codeql} \cite{cosmos-sdk-codeql} consisting of seven queries that identify common vulnerabilities in Cosmos projects. The aforementioned tools with rules tailored to Cosmos SDK projects should, in theory, detect many pitfalls in Cosmos SDK-based blockchains. Nevertheless, they suffer from a lack of accompanying research, resulting in limited understanding regarding their effectiveness. Additionally, many of the live blockchains in the Cosmos network do not use these tools, which serves as another reason to investigate them.

The effectiveness of the tools is composed of three main elements. Firstly, it needs to succeed in actually finding vulnerabilities. Secondly, it should have a high level of precision to minimize the occurrence of false positives, thereby ensuring its practical value. Finally, it needs to be easy to use, maintain and extend.

Our research findings indicate the presence of imprecise rules in cosmos-sdk-codeql, motivating possible enhancements to the existing rule set. This paper focuses on the potential improvement of the current rules to address their shortcomings and enhance their effectiveness, as well as extending the rule set to cover additional common vulnerabilities.

\subsection{Contribution}

First, a corpus comprising 11 Cosmos-based appchains was constructed for conducting subsequent evaluations. The corpus served as the foundation for evaluating the seven cosmos-sdk-codeql queries. The resulting findings were documented, taking into account the number of positives, false positives, noise ratio, and precision.

In the following step, a detailed analysis of the source code of the queries was conducted, with a primary focus on suggesting improvements to enhance their precision. Building upon the findings from the analysis, a refactored query pack was developed, incorporating some of the recommended enhancements. 

To further enhance the capabilities of the query pack, an additional query was introduced, extending the tool's functionality and widening the scope of vulnerability detection.

\section{Background}

\subsection{Cosmos}
\textit{Cosmos} \cite{cosmos-whitepaper} is one of the largest blockchain ecosystems, focused on creating an interconnected network of blockchains, managing over 55 billion dollars in digital assets across more than 240 applications.  To facilitate developers that want to create and connect their blockchain to the cosmos ecosystem, the team has developed the Cosmos SDK. This SDK can quickly bootstrap a blockchain that is based on the Tendermint (prior to v0.47.0) or CometBFT (starting from v0.47.0) consensus engine. These consensus engines communicate with the application using a standardized interface called ABCI (``Application Blockchain Interface'') \cite{abci-0-spec}.

The Cosmos SDK is a modular framework that splits the application into different modules, each with its own separate state. A transaction to the blockchain can affect the state of multiple modules. The SDK introduces a \textit{message} which can be seen as a smaller transaction handled by one specific module. A transaction recorded on the blockchain may thus consist of multiple messages. Additionally, the Cosmos SDK contains a boilerplate implementation of the ABCI in Go, providing the application with a persistent (key, value) store, handling the encoding/decoding and extraction of messages from transactions, routing of these messages to the appropriate module, and committing state changes to the store. The operations that affect the store, and hence the state of the node, will henceforth be called \emph{consensus-critical}. These store updates are required to be deterministic for the blockchain to function correctly, which means that certain common software development practices cannot be used, such as using UIDs to generate unique names or relying on local system time for event timestamps.

\subsection{SAST tools and CodeQL}
A Static Application Security Testing (SAST) tool analyzes the source code of a software application and reports on potential security problems. This analysis can be performed by looking for patterns, data and control flows, security standards, code metrics, and more.

CodeQL \cite{codeql} is a SAST tool that allows developers to detect security vulnerabilities via declarative queries. These custom-written queries make it possible to catch vulnerabilities that are specific to the application. One of the strengths of CodeQL is its ability to analyze code written in various programming languages, including C(++), Go, Java, JavaScript, Python, and more. 

For Go source code, CodeQL is able to reason about the source code in two ways: using an abstract syntax tree (AST), or using data-flow represented in a data-flow graph (DFG). The AST representation allows for reasoning about syntactic properties such as expression types, statement nesting and variable names. The DFG can provide how data flows through the variables at runtime. It also contains information about which function may be invoked by a given call. 

The query language of CodeQL is both declarative and object-oriented \cite{2016avgustinov}. Logical relations can be expressed over the AST and DFG, which simplifies detecting patterns or data flows. A query usually consists of a \texttt{from}, \texttt{where}, and \texttt{select} clause. First, the \texttt{from} clause declares the variables used in the query. Then, the \texttt{where} clause defines any conditions to the variables declared in the \texttt{from} clause. Finally, the \texttt{select} clause specifies the results to display, based on the conditions of the \texttt{where} clause. 

\section{Methodology}

This section covers which Cosmos-based vulnerabilities this paper focuses on. Additionally, we explain how we created the corpus of open-sourced blockchains, which we used to evaluate the CodeQL queries.

\subsection{Vulnerabilities in Cosmos SDK blockchains}

The first step in analyzing the  application security of Cosmos-based blockchains, is figuring out which vulnerabilities exist and how problematic they are. The vulnerabilities of interest, which will be the subject of our investigation, are documented in Table~\ref{tab:rules}. They have been derived from various sources published across the internet. In the table, each vulnerability is accompanied by corresponding references to the sources from which the information has been obtained. Vulnerabilities 1 and 3 are specific to the Cosmos SDK. The others are either specific to Go (2, 4) or can be applied to other languages as well (5, 6, 7, 8). 

{\renewcommand{\arraystretch}{1.1}
\begin{table*}[htp]
\centering
\caption{Vulnerabilities in Cosmos-based blockchains.}
\label{tab:rules}
    \begin{tabularx}{\linewidth}{lX}
        \toprule
        \textbf{Vulnerability ID and name} & \textbf{Explanation} \\\midrule
        1. \{Begin,End\}Block panic & Panics in \texttt{\{Begin,End\}Block} ABCI calls will halt the chain and should never occur \cite{notsosmart} \cite{gosec-issues}. \\
         2. Map iteration & Iterating over a map in Go is non-deterministic and should be avoided when possible. If needed, an iteration over an array of the sorted keys should be used instead \cite{notsosmart} \cite{gosec-issues} \cite{maprange-vuln1}. \\
         3. Hardcoded Bech32 prefix & Bech32 prefixes should not be hardcoded because other modules might use the wrong address values. Instead, they should be configured using the Cosmos SDK's API \cite{gosec-issues} \cite{bech32-vuln1} \cite{bech32-vuln2}. \\
         4. Goroutines & Concurrency using goroutines and select statements are non-deterministic and should be avoided \cite{notsosmart} \cite{gosec-issues}. \\
         5. Floating point arithmetic & Floating point arithmetic can lead to unexpected results and should be avoided when possible. If needed, the operations should be converted into a form that can not result in unexpected behavior \cite{notsosmart} \cite{floating-incident}.\\
        6. System time & System clock times are not synchronized over the network and should never be used to validate or change state \cite{systemtime-vulnerability} \cite{provenance-chain}. \\
        7. Unsafe packages & Some packages contain functions that have unexpected results and are nondeterministic. Such packages include \textit{rand}, \textit{reflect}, \textit{unsafe} and \textit{runtime}, and should be avoided \cite{notsosmart} \cite{gosec-issues}. \\
        8. Platform-dependent types & Using platform-dependent types (e.g. \texttt{int}) can have different results on different architectures and should be avoided \cite{notsosmart}. \\ 
        \bottomrule
    \end{tabularx}
\end{table*}
}

\subsection{Selection of SAST Tools} \label{sec:cosmos-sdk-codeql}

We found three implementations of existing tools that are capable of scanning Cosmos SDK projects for some of our vulnerabilities: CodeQL \cite{codeql} with implemented queries by \citeauthor{cosmos-sdk-codeql} \cite{cosmos-sdk-codeql}, gosec \cite{gosec} with implemented rules by \citeauthor{gosec-cosmos} \cite{gosec-cosmos} and GoLiSa\footnote{https://github.com/lisa-analyzer/go-lisa} \cite{information-flow-analysis-nondeterminism}. We chose to continue on the CodeQL project because of five reasons: 
\begin{enumerate}
    \item CodeQL provides a more declarative interface that makes it easy to write, extend  and maintain rules compared to both gosec and GoLiSa.
    \item GoLiSa only detects non-determinism. However, we want to have the ability to extend the rules to capture vulnerabilities not caused by non-determinism.
    \item The existing query pack cosmos-sdk-codeql consists of more rules than gosec, giving us a larger head start. It also allows for a more elaborate analysis of the results on existing chains.
    \item CodeQL works on multiple languages, while gosec and GoLiSa are specific to Go. Even though the Cosmos SDK is written in Go, the resulting refactored or new CodeQL queries could be more easily understood or ported to other languages.
    \item The Cosmos SDK uses cosmos-sdk-codeql as an automated workflow on GitHub.
\end{enumerate}

\subsection{The corpus of Cosmos SDK-based blockchains}

To analyze the performance of the CodeQL queries, a list of existing chains needs to be tested. At the time of writing, 57 public blockchains were active in the Cosmos ecosystem \cite{mapofzones}. Analyzing the results of CodeQL queries for false positives must be done manually. Hence, we decided to select a subset of these blockchains. It is important to be objective and unbiased when selecting this subset. Therefore, we followed the subsequent restrictions:

\begin{enumerate}
    \item The chain needs to be built with the Cosmos SDK.
    \item The chain needs to be open-source.
    \item The chains need to have distinct use cases; for example, not all of them must be Decentralized Exchanges (DEX).
    \item The chains need to have a variety of ages (based on the date of the first version release).
    \item The chains need to have a variety of sizes (based on the number of transactions submitted to the blockchain between 25 May 2023 and 31 May 2023).
\end{enumerate}

We decided to select 11 blockchains using the following strategy: divide the 57 blockchains into 3 categories based on size. In each of these categories, identify 3 to 5 projects with a distinct use case and with ages spanning at least a year. The chosen blockchains are found in Table~\ref{tab:chains}. Additionally, we included the amount of consensus-critical code function declarations present in each of these blockchains, which gives a better feel for the actual code size. Note that a low consensus-critical code function count does not necessarily indicate a small project. For instance, the Cosmos Hub (Gaia), the center of the network, only has a count of 6 because it mainly uses the prebuilt Cosmos modules, which are considered black-box secure (i.e. not part of the application code).

\renewcommand{\arraystretch}{1.1}
\begin{table*}[htp]
\centering
    \caption{The corpus of the 11 blockchains used in this paper to analyze queries and vulnerabilities. \textit{Date} is the release of the first version. \textit{Total Tx} is the amount of transactions over a span of 7 days, between 25 May 2023 and 31 May 2023. \textit{\#Func} is the total amount of consensus-critical functions.}
        \label{tab:chains}
        \begin{tabular}{lllll}
            \toprule
            \textbf{Name} & \textbf{Use case} & \textbf{Date} & \textbf{Total Tx} & \textbf{\# Func}\\\midrule
            Stride & A liquid staking platform. & 4 Sept. 2022 & 546,690 & 209\\
            Osmosis & The largest interchain decentralized exchange. & 16 June 2021 & 607,470 & 1310\\
            Gaia & The economic center of Cosmos providing IBC security and more. & 13 Aug. 2019 & 289,710 & 6 \\
            Axelar & Web3 integration across multiple chains & 8 Mar. 2021 & 2,944,100 & 2965\\
            Crypto.org & Payment, DeFi and NFTs. & 14 Oct. 2020 & 167,040 & 92\\
            Fetch.ai & Automation of Web3 systems using AI agents. & 1 July 2020 & 264,000 & 2\\
            Regen & Originate and invest in ecological regeneration projects. & 5 June 2019 & 0 & 444\\
            Jackal & A cloud storage solution. & 22 Oct. 2022 & 0 & 257\\
            Medibloc & A patient-centered health data ecosystem. & 26 Aug 2019 & 14,101 & 513\\
            Desmos & A framework to build social media platforms. & 10 Dec. 2019 & 3,202 & 758\\
            Dig& Tokenized real-estate. & 13 Dec. 2021 & 1,370 & 2\\\bottomrule
        \end{tabular}
    \end{table*}

\section{Improving code analysis rules to detect vulnerabilities in Cosmos}

As discussed in Section~\ref{sec:cosmos-sdk-codeql}, CodeQL is the SAST tool that will be used throughout the rest of this paper. \citeauthor{cosmos-sdk-codeql} has already written queries to detect common vulnerabilities in Cosmos SDK blockchains \cite{cosmos-sdk-codeql}, which are referred to as the \textit{cosmos-sdk-codeql} query pack. 
First, we explain why these existing queries are worth refactoring. Next, we cover how we refactored these queries. Lastly, we introduce a new query that detects a vulnerability not covered by any of the discussed SAST tools.

Throughout this paper, we mainly focus on the main script in the projects, i.e., external Cosmos modules are not scanned for vulnerabilities.

\subsection{Analysis of cosmos-sdk-codeql}
The cosmos-sdk-codeql query pack contains queries that are able to detect vulnerabilities 1 to 7 (Table~\ref{tab:rules}). A recurring pattern in most of the queries is the use of a \textit{blacklist} of specific package names to be ignored for further analysis, such as mock packages or packages that store test, simulation, or CLI code. Code in these packages is usually not consensus-critical, which means it does not influence the blockchain state. Therefore, vulnerable code patterns in these packages should be ignored as they are very likely to be false positives. In particular, cosmos-sdk-codeql uses the query \texttt{isIrrelevantPackage} to detect blacklisted packages. This predicate is redefined in every query and has no consistent form: some packages are ignored in a query while they are not ignored in other queries.

The queries detecting vulnerabilities 2 to 7 use the blacklisting approach. Some queries use an exhaustive list of blacklisted packages (e.g. the goroutine query has 18 blacklisted packages), while others only blacklist a few. This creates an inconsistent view of which packages should be ignored. Additionally, this approach only works for certain project structures, as there is no required package naming convention in Cosmos. 

The query to detect \{Begin,End\}Block panics (vulnerability 1) does not use a blacklist. Instead, it finds the \{Begin,End\}Block functions and traverses the DFG to find all functions being called. This proves to be a more consistent approach, as it applies to any project regardless of the package names. On the contrary, this query has two other problems: it flags many duplicate positives and ignores additional true positives present in the same function. 

\subsection{Improving the accuracy of the Cosmos CodeQL rules}
The general approach taken in cosmos-sdk-codeql rules is to scan the entire codebase but use a blacklist to ignore irrelevant code. To improve on this, a \textit{whitelist} approach can be used, i.e. scan only the codebase that contains consensus-critical code. This idea  removes the need for the developer to define which code should be ignored manually and thus can, for example, automatically ignore test files.

In order to ensure reliable vulnerability detection, we should only flag potential vulnerabilities if they occur in consensus-critical code. In the Cosmos SDK version 0.46.12, the following ABCI methods are found to be consensus-critical \cite{abci-0-spec}:

\begin{itemize}
    \item \texttt{BeginBlock}: runs at the start of every block
    \item \texttt{EndBlock}: runs at the end of every block
    \item \texttt{DeliverTx}: runs when a transaction is executed
\end{itemize}

We decided not to include \texttt{InitChain} and \texttt{Commit}, since these are either run once at initialization or not controlled by the application developers. It should be noted that starting from Cosmos SDK version 0.47.0, a new version of ABCI is used, which adds more ABCI methods like \texttt{PrepareProposal} and \texttt{ProcessProposal}. We do not cover this new version here.

Finding consensus-critical code can be made more consistent by utilizing CodeQL classes. We defined a class called \texttt{ConsensusCriticalFuncDecl}, which represents the declaration of all the functions that are used in consensus-critical code. With this class, the false positives of queries can be restricted by requiring that the location of the positive is inside consensus-critical code. Additionally, it is easy to extend this structure. If a new ABCI method becomes available that is considered consensus-critical, a new class can be constructed that covers the nodes used in this method and extends the \texttt{ConsensusCriticalFuncDecl} class.

Using this class definition, we restructured queries 2 to 7. Additionally, we fixed the issue in query 1, discussed in the previous section. The evaluation of the original and refactored queries can be found in Section~\ref{sec:eval}.

\subsection{Identifying additional code vulnerabilities using CodeQL}
Table~\ref{tab:rules} shows how using platform-dependent types can lead to potential vulnerabilities in a blockchain. For example, an integer that is 32 bits wide on one architecture might overflow on a certain operation, while the same operation on a different architecture with an integer width of 64 bits will not. Blockchains should thus not use these types, but none of the discussed tools were able to detect their usage in a Cosmos-based blockchain. Hence, we implemented a new query. It detects the usage of \texttt{int}, \texttt{uint}, and \texttt{uintptr} in consensus-critical code \cite{go-spec}. The results of this new query can be found in the next Section.

\section{Evaluation} \label{sec:eval}
We conducted experiments aimed at evaluating the cosmos-sdk-codeql framework and refined queries. Additionally, we also tested the new query, designed to uncover the use of platform-dependent variables. These evaluations encompassed the projects listed in Table~\ref{tab:chains}. The complete implementation of our experiments, inclusive of query scripts and test data, is accessible through our GitHub repository\footnote{\url{https://github.com/JasperSurmont/cosmos-sdk-codeql}}. 

\subsection{Statistics and results}
The statistics used to evaluate the queries can be seen in Table~\ref{tab:statistics}. Tables~\ref{tab:results-per-rule-old} and \ref{tab:results-per-project-old} display the results of cosmos-sdk-codeql per query and per blockchain, respectively. We see that many queries have very low precision, with only query 1 having 100\%. However, query 1 also has a lot of noise and, as we will see later, misses many true positives. Next, the refactored queries are evaluated, with the results shown in Tables~\ref{tab:results-per-rule-new} and \ref{tab:results-per-project-new}. To put these results in perspective, Tables~\ref{tab:results-per-rule-comp} and \ref{tab:results-per-project-comp} compare the results of cosmos-sdk-codeql and the refactored queries. 

{\renewcommand{\arraystretch}{1.5}
\begin{table}[htb]
    \centering
    \caption{An overview of the statistics used.}
    \label{tab:statistics}
    \begin{tabular}{r|l}
        \textbf{Statistic} & \textbf{Calculation} \\\hline
        Positives (P) & Result of tool \\
        False Positive (FP) & Manual checks \\
        True Positive (TP) & P - FP \\
        Unique True Positive (UTP) & TP - duplicates (manual checks) \\
        Precision  & $\frac{TP}{P}$ \\
        Noise ratio (NR) & $\frac{TP - UTP}{TP}$ \\
        Difference in TP ($\Delta TP$)& $\text{TP}_1 - \text{TP}_2$ \\
    \end{tabular}
\end{table}}

We see that the refactored queries 2 to 7 have a significant reduction in false positives, leading to higher precision. The overall precision of all the blockchains increased by at least 50\% with an average of 80.64\%, and no duplicates were flagged anymore. Additionally, refactored query 1 was able to detect 46 additional true positives. Finally, we see that the reduction in FPs did not have an impact on the found TPs, with two exceptions. The two TPs that were not detected in the refactored queries were found in the initialization of the blockchain. As we were not able to figure out if these could potentially lead to actual vulnerabilities, we decided to leave them in.

\begin{table}[pt]
    \centering
    \caption{The statistics of cosmos-sdk-codeql per query, aggregated over the 11 projects.}
    \label{tab:results-per-rule-old}
    \begin{tabular}{l*{4}{c}}
        \toprule
        Original rules & Positives & UTP & Noise Ratio & Precision \\
        \midrule
        1. \{Begin,End\}Block panic & 45 & 45 & 40\% & 100\% \\
        2. Map iteration & 119 & 5 & 0\% & 4\% \\
        3. Hardcoded Bech32 & 51 & 1 & 1.96\% & 4\% \\
        4. Goroutines & 9 & 0 & N/A & 0\% \\
        5. Floating point & 19 & 0 & N/A & 0\% \\
        6. Local time & 28 & 1 & 0\% & 3.57\% \\
        7. Unsafe packages & 35 & 4 & 0\% & 11.43\% \\
        \bottomrule
    \end{tabular}
    \bigskip

    \caption{The statistics of cosmos-sdk-codeql per project, aggregated over the 7 queries.}
    \label{tab:results-per-project-old}
    \begin{tabular}{l*{4}c}
        \toprule
        Blockchain & Positives & UTP & Noise Ratio & Precision \\\midrule
        Gaia & 1 & 0 & N/A & 0\% \\
        Crypto.org & 3 & 0 & N/A & 0\% \\
        Desmos & 35 & 4 & 43\% & 20\% \\
        MediBloc & 28 & 1 & 50\% & 7.14\% \\
        Stride & 19 & 4 & 20\%  & 26\% \\
        Fetch & 2 & 0 & N/A & 0\% \\
        Dig & 14 & 0 & N/A & 0\% \\
        Regen & 46 & 1 & 0\% & 2.17\% \\
        Jackal & 4 & 1 & 50\% & 50\% \\
        Osmosis & 58 & 7 & 36\% & 18.97\% \\
        Axelar & 97 & 20 & 31\% & 29.90\% \\
        \bottomrule 
    \end{tabular}
\end{table}
\begin{table}[pt]
    \centering
    \caption{The statistics of the refactored queries, aggregated over the 11 projects.}
    \label{tab:results-per-rule-new}
    \begin{tabular}{l*{4}{c}}
        \toprule
        Refactored / new queries & Positives & UTP & NR & Precision \\
        \midrule
        1. \{Begin,End\}Block panic & 91 & 91 & 0\% & 100\% \\
        2. Map iteration & 13 & 5 & 0\% & 38\% \\
        3. Hardcoded Bech32 & 0 & 0 & N/A & N/A \\
        4. Goroutines & 0 & 0 & N/A & N/A \\
        5. Floating point & 2 & 0 & N/A & 0\% \\
        6. Local time & 0 & 0 & N/A & N/A \\
        7. Unsafe packages & 5 & 4 & 0\% & 80\%  \\
        8. Platform dependent types & 44 & 35 & 0\% & 79.54\% \\
        \bottomrule
    \end{tabular}
    \bigskip

    \caption{The statistics of the refactored queries per project, aggregated over the 7 queries.}
    \label{tab:results-per-project-new}
    \begin{tabular}{l*{4}c}
        \toprule
        Blockchain & Positives & UTP & Noise Ratio & Precision \\\midrule
        Gaia & 0 & 0 & N/A & N/A \\
        Crypto.org & 0 & 0 & N/A & N/A \\
        Desmos & 8 & 5 & 0\% & 62.5\% \\
        MediBloc & 3 & 3 & 0\% & 100\% \\
        Stride & 5 & 5 & 0\%  & 100\% \\
        Fetch & 0 & 0 & N/A & N/A \\
        Dig & 0 & 0 & N/A & N/A \\
        Regen & 5 & 5 & 0\% & 100\% \\
        Jackal & 6 & 6 & 0\% & 100\% \\
        Osmosis & 101 & 86 & 0\% & 85.15\% \\
        Axelar & 33 & 31 &  0\% & 93.94\% \\
        \bottomrule 
    \end{tabular}
\end{table}
\begin{table*}[pt]
    \centering
    \caption{The comparison of cosmos-sdk-codeql and refactored queries per query, aggregated over the 11 projects. $FP_c$ / $UTP_c$ are the FPs / UTPs of cosmos-sdk-codeql, $FP_r$ / $UTP_r$ are the FPs / UTPs of the refactored queries. For $\Delta FP$ and $\Delta NR$, lower is better.}
    \label{tab:results-per-rule-comp}
    \begin{tabular}{l*{8}{c}}
    \toprule
        Query & $FP_{c}$ & $FP_{r}$ & $UTP_{c}$ & $UTP_{r}$ & $\Delta FP$ & $\Delta UTP$ & $\Delta NR$ & $\Delta Prec$ \\
        \midrule
        \shortstack[l]{1. \{Begin,End\}Block \\ panic} & 0 & 0 & 45 & 94 & 0 & 46 & -40\% & 0\% \\
        2. Map iteration & 115 & 8 & 5 & 5 & -106 & 0 & 0\% & 34.26\% \\
        3. Hardcoded Bech32 & 49 & 0 & 1 & 0 & -49 & -1 & N/A & N/A \\
        4. Goroutines & 9 & 0 & 0 & 0 & -9 & 0 & N/A & N/A \\
        5. Floating point & 19 & 2 & 0 & 0 & -17 & 0 & N/A & N/A \\
        6. Local time & 27 & 0 & 1 & 0 & -27 & -1 & N/A & N/A \\
        7. Unsafe packages & 31 & 1 & 4 & 4 & -30 & 0 & 0\% & 68.57\% \\
    \bottomrule
    \end{tabular}
    \bigskip
    \caption{The comparison of cosmos-sdk-codeql and refactored queries per project, aggregated over the 7 queries. $FP_c$ / $UTP_c$ are the FPs / UTPs of cosmos-sdk-codeql, $FP_r$ / $UTP_r$ are the FPs / UTPs of the refactored queries. For $\Delta FP$ and $\Delta NR$, lower is better.}
    \label{tab:results-per-project-comp}
    \begin{tabular}{l*{8}{c}}
        \toprule
        Project & $FP_{c}$ & $FP_{r}$ & $UTP_{c}$ & $UTP_{r}$ & $\Delta FP$ & $\Delta UTP$ & $\Delta NR$ & $\Delta Prec$ \\
        \midrule
        Gaia & 1 & 0 & 0 & 0 & -1 & 0 & N/A & 100\% \\
        Crypto.org & 3 & 0 & 0 & 0 & -3 & 0 & N/A & 100\% \\
        Desmos & 28 & 3 & 7 & 5 & -25 & 1 & -43\% & 42.5\% \\
        MediBloc & 26 & 0 & 1 & 3 & -26 & 2 & -50\% & 92.86\% \\
        Stride & 14 & 0 & 5 & 5 & -14 & 1 & -20\% & 73.68\% \\
        Fetch & 2 & 0 & 0 & 0 &-2 & 0 & N/A & 100\% \\
        Dig & 14 & 0 & 0 & 0 &-14 & 0 & N/A & 100\% \\
        Regen & 45 & 0 & 1 & 5 & -45 & 4 & 0\% & 97.83\% \\
        Jackal & 2 & 0 & 1 & 6 & -2 & 5 & -50\% & 50\%\\
        Osmosis & 47 & 15 & 7 & 79 & -32 & 79 & -36\% & 66.18\% \\
        Axelar & 68 & 2 & 20 & 31 & -66 & 11 & -31\% & 64.04\% \\
        \bottomrule
    \end{tabular}
\end{table*}

\subsection{Effectiveness and usefulness of the analyzed queries}
The quantitative numbers of the results might be misleading. As such, we discuss the results of the refactored queries to identify their effectiveness and usefulness.

\paragraph{Hardcoded Bech32, Goroutines, Floating point, Local time and Unsafe packages} These queries  resulted in 0 or very few positives (and hence 0 or little TPs) and might seem redundant or not worth incorporating and maintaining. However, a low positive count does not necessarily mean that the query is useless. The repositories that were analyzed are all open-source and backed by a team of developers. Hence, obvious vulnerabilities are likely to be noticed and fixed quickly. Additionally, the vulnerabilities that these queries detect have a fixed form and are easy to spot. For example, using \texttt{time.Now()} or a goroutine is likely spotted by an experienced developer.

It is important to maintain focus on the scope of Cosmos SDK. This SDK is not only used for public blockchains backed by a team but also for learners, individuals, and private blockchains. Inexperienced developers might not have the ability to detect these vulnerabilities as easily, which is where the queries would come in handy. 

\paragraph{\{Begin,End\}Block panic} This query has a lot of true positives, with the majority of these positives centralized in the Osmosis and Axelar blockchains. The reason for this centralization is that the Cosmos community does not agree on whether these functions are allowed to panic or not. One of the maintainers of the Cosmos SDK stated that: \textit{"There should be no unanticipated panics in these methods, meaning that sometimes you will want to panic if things have gone horribly wrong"} \cite{cosmos-panics-discussion}. On the other hand, multiple sources claim that panic should never happen \cite{notsosmart} \cite{gosec-issues} \cite{cosmos-sdk-codeql}, and lots of blockchains (including Gaia, maintained by the Cosmos team) avoid panics in these functions.

Regardless of the stance of the community, \textit{unexpected} panics should never occur. The query is thus still very useful for detecting panic flows that developers do not anticipate. In the future, if the community decides that expected panics are acceptable, there should be a way to flag these expected panics to remove the FPs.

\paragraph{Map iteration} Out of all the refactored queries, the map iteration query has the lowest precision. This is mainly due to the complexity of detecting whether operations inside the loop actually affect the state or not. In the cosmos-sdk-codeql repository, it has even been suggested to remove this query entirely since it is violated too often because of the amount of FPs \cite{removal-of-map-iter}. 

We argue that this query is important and should thus not be removed. First, the accuracy of this query can be increased. For example, it could take a similar approach to what the GoLiSa tool does \cite{information-flow-analysis-nondeterminism}, which acquires a high precision rate. Second, vulnerabilities caused by map iterations were the source of multiple important incidents \cite{thor-halt} \cite{maprange-vuln1}, so completely removing the query would fail to catch these vulnerabilities. Third, inexperienced developers are likely not aware of the non-determinism that map iteration in Go causes. A static check, like this query, can thus assist the development process instead of spending time on dynamic code debugging.

\paragraph{Platform dependent types} This query has a quite high precision of almost 80\%, with 35 TPs found in the analyzed blockchains. This is a bit misleading, as in many cases, the odds of the value of the variable showing non-determinism are extremely low. Indeed, asserting that the discovered vulnerabilities are always dangerous is incorrect. Nevertheless, replacing the platform-dependent types with independent types should not be an issue and completely removes the possible vulnerabilities that they cause. 

\subsection{Overall state of the security of blockchains using the Cosmos SDK}

We have seen research and multiple tools that try to detect vulnerabilities in Cosmos SDK applications. Of the 11 chains that we analyzed, 4 used such a tool in the GitHub workflow (all 4 used CodeQL with the cosmos-sdk-codeql query pack). The remaining 7 do not use any tool with rules specifically for the Cosmos SDK in their CI pipeline. 

One reason for the low tool usage can be attributed to the amount of false positives. For example, running cosmos-sdk-codeql on Osmosis provides 56 positives, of which only 11 were found to be true positives. Of these 11 TPs, all of them were panics in the \texttt{\{Begin,End\}Block} functions, which they might not consider as a vulnerability. Having too many FPs reduces the attractiveness of the tools and will decrease the frequency with which they are used.

Another possible reason is that these tools are not advertised in the Cosmos SDK documentation or tutorials, making it not straightforward for new developers to discover these tools.

The Cosmos SDK is designed to increase the robustness and speed of the development while decreasing the amount of skill required to build an application-specific blockchain. However, concrete best practices that increase the security of the blockchain are not easy to find and require extensive research. This is a bit contradictory to what the Cosmos SDK aims to achieve. It would be beneficial if an explanation of these best practices were easily accessible, or if a tool such as CodeQL with appropriate queries was referenced in the documentation.

\section{Conclusion}
Application-specific blockchains are an emerging class of blockchains that use general-purpose languages to define the blockchain logic, offering the promise of flexibility and speed at the risk of an increased attack surface, particularly non-deterministic execution. SAST tools can help identify such issues early during development.

The Cosmos network is currently one of the largest blockchain networks, managing over 55 billion USD in digital assets. The Cosmos SDK is a framework that helps developers build application-specific blockchains. To improve security during development, SAST tools can detect  vulnerabilities that pose potential security risks to the blockchain, such as cosmos-sdk-codeql, which is, for instance, present in the Cosmos SDK's automated GitHub workflow. However, the overall underutilization of these tools across numerous blockchains prompted an investigation.

After analyzing the cosmos-sdk-codeql query pack on our corpus of appchains, we found that many queries had very low precision, with 6 out of 7 queries having a precision value between 0\% and 11.43\% across our corpus. These low precision values mainly result from an inconsistent strategy to filter out false positives: 64\% of all false positives originated from 3 out of the 11 blockchains (Osmosis, Axelar, and Regen). Low precision rates of the queries result in a diminished adoption rate, as developers are likely reluctant to cope with many false positives. 

We refactored the queries to have fewer false positives by detecting, with higher precision, the code that is "consensus-critical" (i.e., part of the blockchain transaction logic), and then including results only from such consensus-critical code. As a result, the precision increase of the 11 blockchains was between 42.5\% and 100\%, with an average rise of 80.64\%. The number of false positives per query decreased by 34 false positives out of an average of 44 total positives.

We further extended the ruleset by implementing an additional rule to detect the use of platform-dependent types. The rule generated 35 true positives across the corpus. However, we could not confirm if any of these true positives are a potential vulnerability.

While no new confirmed vulnerabilities were identified among the 11 tested blockchains, the significant reduction in false positives makes the tool more effective across all Cosmos-based appchains. We hope this result may inspire more blockchain developers to consider using SAST tools as part of their software development cycle to detect potential issues earlier.

\bibliographystyle{IEEEtranN}
\bibliography{mybibliography.bib}

\vspace{12pt}

\end{document}